\documentclass[a4paper,12pt]{article}
\usepackage{epsfig}
\usepackage{wrapfig}
\usepackage{sidecap}
\usepackage{amssymb}
\begin{document}
\thispagestyle{empty}

\newcommand{\etal}  {{\it{et al.}}}  
\def\Journal#1#2#3#4{{#1} {\bf #2}, #3 (#4)}
\def\PRD{Phys.\ Rev.\ D}
\def\NIMA{Nucl.\ Instrum.\ Methods A}
\def\PRL{Phys.\ Rev.\ Lett.\ }
\def\PLB{Phys.\ Lett.\ B}
\def\EPJ{Eur.\ Phys.\ J}
\def\IEEETNS{IEEE Trans.\ Nucl.\ Sci.\ }
\def\CPCD{Comput.\ Phys.\ Commun.\ }

\bigskip

{\bf
\begin{center}
 \textbf{\large {Bose-Einstein correlations of two identical particles through their first measurements
 by CMS at the LHC} }
\end{center}
}


\begin{center}
{\large G.A. Kozlov  }
\end{center}


\begin{center}
\noindent
 {
 Bogolyubov Laboratory of Theoretical Physics\\
 Joint Institute for Nuclear Research,\\
 Joliot-Curie st., 6 Dubna\\
 141980 Moscow region, Russia
 }
\end{center}

\begin {abstract}
\noindent
{
We look at the new two-particle correlation function that
can provide the space-time information about the source of two charged pions.
This function clearly reflects the chaotic and incoherent nature of the particle emission source 
disturbed by external forces (fields).
We argue that such an investigation could probe the size and the temperature of the expanded 
source where two pions are emitted in thermal environment.}
\end {abstract}



\begin{center}
\noindent
\end{center}



1. The Large Hadron Collider (LHC) at CERN is at the stage to provide particle physicists with treasure of data. 
These data allowed a precise measurement of many important parameters of modern particle physics
in order to test their consistency in particular to some theoretical models.
The phenomenon of Bose-Einstein correlations (BEC) is clear and undeniable
part of new physics, complicating the quantum statistical description of multi-particle final states at finite temperatures.

Historically, the investigations of particle correlations have been concentrated on pion pairs correlations, however have also been
applied to more heavy hadrons, quarks, protons, and even gauge bosons, photons, $Z$-bosons etc. (see, e.g., [1-4] 
and the references therein).

Recently, the first measurements of BEC in proton-proton collisions at $\sqrt {s}$ = 0.9 and 2.36 TeV
center-of-mass energies recorded by the CMS experiment at the LHC have been carried out [5]. In particular,
the size of the particle emission region $r$ is seen to increase significantly with the particle
multiplicity $N$. However, the intercept (or chaoticity) parameter $\lambda$ has
the non monotonic behavior with $N$. The dependence of $r$ and $\lambda$ on both  $N$ and $\sqrt {s}$,
disputed for a long time ( see, e.g., [6]), is now very clearly observed in $pp$ collisions [5].

This letter gives the extension of BEC phenomenology based of the first principles within the Quantum Field Theory
at finite temperature, that opens the new horizon in experimental study of BEC itself.
One of the main parameters is the temperature of the particle (emitting) source under
the random external forces (fields) influence [7-12].


By studying BEC of identical particles, it is possible
to determine the time scale and spatial region over which particles do not
have the interactions. Such a surface is called as decoupling one.
In fact, for an evolving system such as $p p$ collisions, it is
not really a surface, since at each time there is a spread out
surface due to fluctuations in the last interactions, and the shape
of this surface evolve even in time. The particle source is not
approximately constant because of energy-momentum conservation
constraint.

In this letter, we make an attempt to demonstrate that the problem
of properties of the genuine interactions can be explored using
experimental data collected at the LHC.
These data can be analyzed through the compared
measures of some inclusive distributions and final state
correlations.

2. A pair of identical bosons with momenta $p_{1}$ and $p_{2}$ and the
mass $m$ produced incoherently from an extended source will
have an enhanced probability $C_{2}(p_{1},p_{2})=
N_{12}(p_{1},p_{2})/[N_{1}(p_{1})\cdot N_{2}(p_{2})]$ to be measured
in terms of differential cross section $\sigma$, where
 $$N_{12}(p_{1},p_{2})=\frac{1}{\sigma}\frac{d^{2}\sigma}{d\Omega_{1}\,d\Omega_{2}} $$
to be found close in 4-momentum space $\Re_{4}$ when detected
simultaneously, as compared to if they are detected separately with
$$ N_{i}(p_{i})=\frac{1}{\sigma}\frac{d\sigma}{d\Omega_{i}}, \,\,\,
 d\Omega_{i}=\frac{d^{3}\vec p_{i}}{(2\pi)^{3}\,2E_{p_{i}}}, \,\,
 E_{p_{i}}=\sqrt {\vec p_{i}^{2}+m^{2}},\,\,\,
 i = 1, 2. $$
In an experiment, one can account the inclusive density $\rho_{2}(p_{1}, p_{2})$
which describes the distribution of two particles in $\Omega$ (the sub-volume of the phase space)
irrespective of the presence of any other particles
$$ \rho_{2}(p_{1}, p_{2}) = \frac{1}{2!}\frac{1}{n_{events}}\,\frac{d^{2} n_{2}}{dp_{1}\,dp_{2}},$$
where $n_{2}$ is the number of particles counted in a phase space domain
$(p_{1} + dp_{1},p_{2} + dp_{2})$. The multiplicity $N$ normalization stands as
$$\int_{\Omega} \rho_{2}(p_{1}, p_{2})\,dp_{1}\,dp_{2} = \langle N (N-\delta_{12})\rangle, $$
where $\langle N\rangle$ is the averaged number of produced particles. Here, $\delta_{12} =0$
for different particles, while $\delta_{12} =1$ in case of identical ones (coming from the same event).

On the other hand, the following relation can be used to retrieve the BEC function
$C_2(Q)$:
\begin{equation}
\label{e23}
 C_2(Q) = \frac{N(Q)}{N^{ref}(Q)},
\end{equation}
where $N(Q)$, in general case, is the number of particle pairs in BEC pattern with
\begin{equation}
\label{e24}
 Q = \sqrt {-(p_1-p_2)_{\mu}\cdot (p_1-p_2)^{\mu}}= \sqrt{M^{2} - 4\,m^{2}}.
\end{equation}
In definitions  (\ref{e23}) and  (\ref{e24}),  $N^{ref}$ is the number of
particle pairs without BEC and
$p_{\mu_{i}}= (\omega_{i}, \vec p_{i})$ are four-momenta of produced particles $(i = 1,\ 2)$;
$M = \sqrt {(p_1+p_2)^{2}_{\mu}}$ is the invariant mass of the pair of particles.

An essential problem is the estimation of the
reference distribution $N^{ref}(Q)$ in (\ref{e23}). If there are other
correlations besides the Bose-Einstein effect, $N^{ref}(Q)$ should be
replaced by a reference sample corresponding to the two-particle distribution
in a geometry without BEC. Thus, the expression (\ref{e23}) represents the ratio
between the number of particles pairs $N(Q)$ in the real world and the reference sample
$N^{ref}(Q)$ in the imaginary world. Note, that  $N^{ref}(Q)$ can not be directly
observed in an experiment. Different methods are usually applied for the construction
of $N^{ref}(Q)$ [1], however all of them have strong restrictions. One of the preferable
methods is to construct it directly from data, for example, to use the particles pairs from 
different (mixed) events.

It is commonly assumed that the maximum magnitude of the function $C_2(Q)$ is 2
for $\vec p_{1} = \vec p_{2}$ if no any distortion and final state interactions are
taking into account.

In general, the shape of $C_2(Q)$ is model dependent. Its most simple form is [13,14]:
\begin{equation}
 \label{e25}
 C_2(Q)=C_0\cdot \left [1+\lambda\exp {(- Q\cdot r)^{\kappa}}\right] \cdot (1+\varepsilon Q) ,
\end{equation}
where $C_0$ is the normalization factor, $\lambda$ is the
chaoticity strength factor, meaning $\lambda =1$ for fully
incoherent and $\lambda =0$ for fully coherent sources; the
symbol $r$ is often called as the "correlation radius", and assumed to be
spherical in this parameterization.  The linear term in (\ref{e25}) accounts
for long-range correlations outside the region of BEC.
For the experimental data fitting it is often used either $\kappa = 1$ or  $\kappa = 2$.

Note that the distribution of bosons
can be either far from isotropic, usually concentrated in some directions, or
almost isotropic, and what is important that in both cases the particles
are under the random chaotic influence or interactions caused by other fields in the thermal
bath. In the parameterization (\ref{e25}) all of these issues
should be  embedded in $\lambda$.

To enlarge the quantum pattern of particle production process at finite temperature,
we use $C_2(Q)$ obtained in [2,3] in the following functional form:
\begin{equation}
\label{e26}
C_2(p_{1},p_{2}) \simeq \xi(N) \left \{1 + \lambda_{1}(\beta)\,e^{-\Delta_{p\Re}}
\left [1+\lambda_{2}(\beta)\,e^{ +\Delta_{p\Re}/2}\right ]\right\} ,
\end{equation}
where 
$\Delta_{p\Re} =  (p_{1} - p_{2})^{\mu}\,\Re_{\mu\nu}\,(p_{1} - p_{2})^{\nu}$  
is the smearing smooth dimensionless generalized function. $\Re_{\mu\nu}$ is the
nonlocal structure tensor of the space-time size, and it defines the domain of emitted particles.
$\xi(N)$ depends on the multiplicity $N$ as $ \xi(N)= \langle{N (N-1)}\rangle/\langle N\rangle^2$.
The functions $ \lambda_{1} (\beta)$ and  $ \lambda_{2} (\beta)$ are the measures
of the strength of BEC between two particles: $\lambda_{1} (\beta)=
\gamma(\omega,\beta)/(1+\alpha)^{2}$, and the correction to the coherence
function in the brackets of (\ref{e26}) is
$\lambda_{2}(\beta) = 2\,\alpha/\sqrt{\gamma(\omega,\beta)}$.
 The function $\gamma (\omega,\beta)$ calls for quantum thermal features of BEC
pattern and is defined as
\begin{equation}
\label{e27}
 \gamma (\omega,\beta) \equiv \gamma (n)  = \frac{{n^2 (\bar \omega )}}{{n(\omega )\ n(\omega
 ')}} ,\ \
 n(\omega ) \equiv  n(\omega ,\beta ) =
 \frac{1}{{e^{\omega \beta} - 1 }} ,\ \
 \bar\omega  = \frac{{\omega  + \omega '}}{2} ,
\end{equation}
where $n(\omega,\beta )$ is the mean value of quantum numbers for Bose-Einstein
(statistics) particles with the energy $\omega$ in the thermal bath
with statistical equilibrium at the temperature $T= 1/\beta$. The
following condition $\sum_{f} n_{f}(\omega,\beta) = N$ is evident,
where the discrete index $f$ stands for the one-particle state $f$.

The important function $\alpha = \alpha (\beta, m)$ entering $\lambda_{1}$ and $\lambda_{2}$ 
in (\ref{e26}), the measure of chaoticity,
summarizes our knowledge of other than space-time characteristics of the particle
emitting source, and it varies from $0$ to $\infty$.

In terms of time-like $L_{0}$, longitudinal $L_{l}$ and transverse $L_{T}$ components of the
space-time size $\sqrt {L^{2}_{\mu}}$, the distribution $\Delta_{p\Re}$ looks like
\begin{equation}
\label{e266}
\Delta_{p\Re}\rightarrow \Delta_{pL} = (\Delta p^{0})^{2}\,L_{0}^{2} +
(\Delta p^{l})^{2}\,L_{l}^{2} + (\Delta p^{T})^{2}\,L_{T}^{2},
\end{equation}
where $L_{0}$  is treated as the measure of the particle emission time,
or even it represents the interaction strength of outgoing particles.

 Hence, we have introduced a new parameter $\sqrt {L^{2}_{\mu}}$,  which defines the region
 of non-vanishing particle density with the space-time extension of the particle emission source.
 Formula (\ref{e26}) must be understood in the sense that $ \exp (-\Delta_{p\Re})$ is a distribution that
 in the limit $L\rightarrow\infty $ strictly becomes a $\delta$ - function. For practical using with
 ignoring the energy-momentum dependence of $\alpha$, one has:
\begin{equation}
\label{e26666}
C_2(Q) \simeq \xi(N) \left \{1 + \lambda_{1}(\beta)\,e^{-Q^2 L^{2}}
\left [1+\lambda_{2}(\beta)\,e^{+Q^{2} L^{2}/2}\right ]\right\}.
\end{equation}
The parameter $L = L(\beta, m)$ in formula (\ref{e26666}) is the measure of the space-time
overlap between two particles, and the physical meaning of $L$ depends on the fitting of
$C_{2}(Q)$-function.
$L$ can be defined through the evaluation of the root-mean-squared momentum $Q_{rms}$ as:
$$Q_{rms}^{2} (\beta) =\langle \vec Q^{2}\rangle =
\frac{\int_{0}^{\infty} d\vert\vec Q\vert\, \vec Q^{2}\,\left [\tilde C_{2}(Q,\beta) -1 \right ]}
{\int_{0}^{\infty} d\vert\vec Q\vert\,\left [\tilde C_{2}(Q,\beta) -1 \right ]},\,\,\,\,
\tilde C_{2}(Q,\beta) =\frac{C_{2}(Q,\beta)}{\xi (N)},$$
where $L$ and $Q_{rms}(\beta)$ are related to each other by means of
$$ L(\beta, m) = \left [\frac{3}{2}\left (1+\frac{1}{1+\frac{1}{4\,\alpha(\beta, m)}\,\sqrt \frac{{\gamma (n)}}{2}}\right )
\right ]^{1/2}\frac{1}{Q_{rms}(\beta)}. $$
The following restriction $\sqrt {3/2} < (L\cdot Q_{rms}(\beta)) < \sqrt{3}$ is evident, where
the lower bound satisfies to the case $\alpha\rightarrow 0$ (no any distortion in the particle production
domain), while the upper limit is given by the very strong influence of chaotic external fields (forces),
$\alpha\rightarrow \infty$. The result is rather stable in the wide range of variation of $\alpha$.

3. We have to emphasize that there are two different scale parameters when the BEC are explored [2,3].
One of them is the correlation radius $r$ introduced
in (\ref{e25}). In fact, the latter gives the "pure" size of the particle emission
source without the influence of the distortion and interaction forces coming from
other fields. The other (scale) parameter is the
scale $ L = L_{st}$ of the production particle domain where the stochastic,
chaotic distortion due to environment is enforced.
This stochastic scale carries  the dependence of the particle mass,
the $\alpha$-coherence degree and what is very important - the temperature dependence.

The question arises: how can BEC be used to determine the effective stochastic scale $L_{st}$
and, perhaps, the phase transition?

At low temperatures ($T < n\sqrt {m^{2} + \mu^{2}}, \,n = 1,2,3, ...)$, in the case of
 two charged pions with the mass $m$, $L_{st}$ has the form ($\mu$ is the chemical potential):
\begin{equation}
\label{e31}
L_{st} (T)\simeq {\left [\frac{e^{\sqrt {m^{2} + \mu^{2}}/T}}
{ 3\,\alpha(N)\,\left (\Delta\epsilon_{p}\right )^{2}\,
(m^{2} + \mu^{2})^{3/4}\, (T/2\,\pi)^{3/2}\, \left
(1+\frac{15}{8}\frac{T}{\sqrt {m^{2} + \mu^{2}}}\right )}\right ]}^{\frac{1}{5}},
\end{equation}
where
$$ \left(\Delta\epsilon_{p}\right )^{2}\simeq m^{2}{\left \vert 1 - \sqrt{2- \frac{2\,E_{\mu}}{m} +
\left (\frac{m_{\mu}}{m}\right )^{2}}\right\vert }^{2}, $$
and $E_{\mu}\simeq $ 4.12 MeV been the energy of the muon with the mass $m_{\mu}$ in the
decay $\pi^{\pm}\rightarrow \mu^{\pm}\,\nu $.

It turns out that the scale $L_{st}$ defines the range of stochastic forces.
This effect is given by $\alpha (N)$-coherence degree which can be estimated from
the experiment within the function $C_{2}(Q)$ as $Q$ close to zero, $C_{2}(0)$,
at fixed value of mean multiplicity $\langle N\rangle$:
\begin{equation}
 \label{e33}
\alpha (N) = \frac{1 + \gamma^{1/2}(n) -\tilde C_{2}(0) + \gamma^{1/4}(n)
\sqrt {\tilde C_{2}(0) [\gamma^{1/2}(n) -2 ] + 2}}{\tilde C_{2}(0)-1}.
\end{equation}
\begin{wrapfigure}{R}{0.5\textwidth}
\renewcommand{\figurename}{Fig.}
  \centering
    \includegraphics[width=0.48\textwidth, height = 85mm]{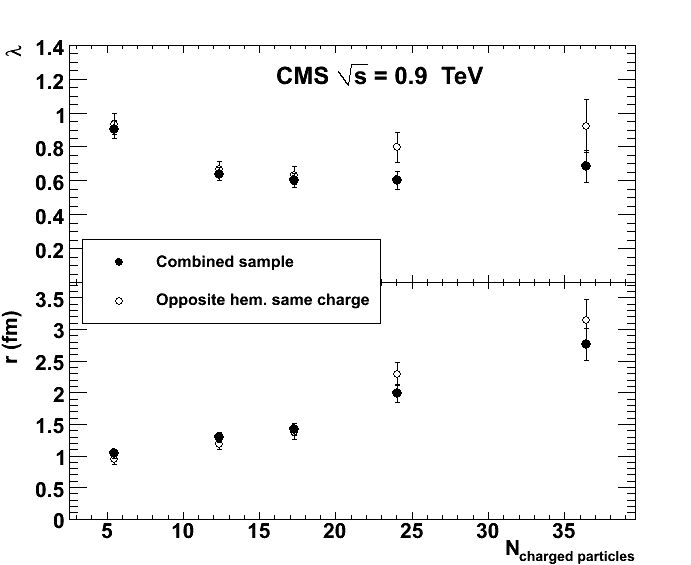}
    \caption{{ \it Values of the $\lambda$ (top) and r (bottom) parameters as a function of the charged-particle multiplicity in the event for combined (dots) and opposite-hemisphere, same-charge (open circles) reference samples, at 0.9 TeV. The errors shown are statistical only. The points are placed on the horizontal scale at the average of the multiplicity distribution in the corresponding bin (borrowed from [5]).}}
  \label{fig:secondgraph}
\end{wrapfigure}
The upper limit of $C_{2}(0)$ depends on $\langle N\rangle$ as well as the quantum thermal features
of BEC pattern given by $\gamma (n)$: $C_{2}(0) < \xi (N) [1 - \gamma^{1/2}(n)/2 ]^{-1}$. This
upper limit is restricted by the maximal value of 2 in the ideal case as $\langle N\rangle \rightarrow\infty$
and $\gamma (n) =1$.


Note, that for $C_{2}(Q)$ - function (\ref{e26}), the limit $\alpha\rightarrow\infty$
yields for fully coherent sources with small $\langle N\rangle$,
while $\alpha\rightarrow 0$ case stands for fully chaotic (incoherent) sources as
$\langle N\rangle \rightarrow\infty$.
Actually, the increasing of $T$ leads to squeezing of the domain of
stochastic forces influence, and $L_{st}(T=T_{0})= r$ at some effective temperature $T_{0}$.
The higher temperatures, $T > T_{0}$, satisfy to more squeezing effect and at the critical temperature
$T_{c}$ the scale $L_{st}(T=T_{c})$ takes its minimal value.
Obviously, $T_{c}\sim O(200~MeV)$ defines the phase transition where
the chiral symmetry restoration will occur.
Since in this phase all the masses tend to zero  and $\alpha\rightarrow 0$, one
should expect the sharp expansion of the region with $L_{st}(T>T_{c})\rightarrow \infty$.

Looking through the formulas  (\ref{e31}) and (\ref{e33}) one can easily find the non-linear
increasing of the linear size of the emitter source with the charged-particle multiplicity.
This dependence is clearly seen in the CMS experiment (see  the bottom panel in Fig. 1).

In terms of the chaoticity parameter $\lambda$ which is related to $\alpha (N)$ through
$$\alpha (N) \simeq \frac{1-\lambda + \sqrt {1-\lambda}}{\lambda},\,\, 0 < \lambda \leq 1 , $$
we expect the non-monotonic behavior of $\lambda (N)$ with $N$. This pattern has also been
observed by CMS (see the upper panel in Fig. 1). The following behavior is evident:
$\lim_{N\rightarrow\infty} \alpha (N)\rightarrow 0$ which tends to
$\lim_{N\rightarrow\infty} r (T_{0})\rightarrow \infty$.

The qualitative relation between $r$ and $L_{st}$ is the only one we
can emphasize in order to explain the mass and temperature dependencies of the source size.
The dependence of the chaoticity $\alpha (N)$ on the minimum scale cut $L_{st}$
and $Q_{rms}$ can be used to define the fit region for different transverse momentum $p_{T}$. 
Such a minimum cut on $L_{st}$ introduces a lower cutoff on $r$.

Apparatus or analysis effects which may result in the sensitivity of external random forces
influence, may be investigated by studying the dependence of the correlation functions on
the stochastic scale $L_{st}$. This effect is expected to contribute strongly at small $L_{st}$
(or large $\alpha$ and $T$).

4. \textbf{To summarize:} the theoretical model for two-particle Bose-Einstein correlation
function in case of two charged pion pairs in $pp$ collisions is carried
out for the first time. One of the main reasons to study BEC at finite temperature
is the possibility to determine the precision with which the source size
parameter and the strength chaoticity parameter(s) can be measured at particle colliders.

Such investigations provide an opportunity
for probing the temperature of the particle production source and the
details of the external forces chaotic influence.

In this letter, we faced to $C_{2}(Q,\beta)$ - function in which the
contribution of $N$, $T$, $\alpha$ are presented.
This differs from the methods already used (see, e.g., [1,6]) which resemble the traditional way of BEC study,
however any qualitative interpretations of $r$, $\lambda$, $\epsilon$ have not been  clarified.

We find that the stochastic scale $L_{st}$ decreases with increasing $T$ slowly at low
temperatures, and it decreases rather abruptly when the critical temperature is approached.


Actually, the experimental measuring of $r$ (in $fm$) ensures the precise estimation
of the effective temperature $T_{0}$ which is one of the main thermal characters
in the particle pair emitter source at given $\alpha (N)$ fixed by both  $C_{2}(Q=0)$ and $\langle N\rangle$.
$T_{0}$ is the true temperature in the region of multiparticle production with dimension
$r = L_{st}$, because at this temperature it is exactly the creation of two
quanta (pions) occurred in $pp$ collisions, and these particles obey the criterion of BEC.

Using the CMS data [5], we point out that the effects of random chaotic forces/fields and the temperature $T_{0}$ 
are decreasing with $\sqrt {s}$, namely, $\alpha (N)$ =1.58 and $ T_{0}$ = 107 MeV at $\sqrt {s}$ = 0.9 TeV, while
$\alpha (N)$ =1.38 and  $ T_{0}$ = 88 MeV at $\sqrt {s}$ = 2.36 TeV.
\textbf{Acknowledgements:} the author would like to thank I. Gorbunov for the help in the preparation of the manuscript. 
\begin{center}
REFERENCES

\end{center}

1. R.M. Weiner, Phys. Rep. 327 (2000) 249.\\


 2. G.A. Kozlov, Phys. Nucl. Part. Lett. 6 (2009) 162.\\

 3. G.A. Kozlov, Phys. Nucl. Part. Lett. 6 (2009) 177.\\

 4. G.A. Kozlov, arXiv:1005.0057 [hep-ph]. \\

 5. V. Khachatryan et al. (CMS Collaboration), Phys. Rev. Lett. 105 (2010) 032001.\\

 6. W. Kittel and E.A. De Wolf, Soft Multihadron Dynamics
 (World Scientific, Singapore, 2005).\\

 7. G.A. Kozlov, Phys. Rev. C58 (1998) 1188.\\

 8. G.A. Kozlov, J. Math.  Phys. 42 (2001) 4749.\\

 9. G.A. Kozlov, New J. of Physics 4 (2002) 23.\\

 10. G.A. Kozlov, O.V. Utyuzh and G. Wilk, Phys. Rev. C68 (2003) 024901.\\

 11. G.A. Kozlov, J. Elem. Part. Phys. Atom. Nucl. 36 (2005) 108.\\

 12. G.A. Kozlov, O.V. Utyuzh, G. Wilk, W. Wlodarczyk, Phys. of Atomic Nucl.
  71 (2008) 1502.\\



13. G. Goldhaber et al., Phys. Rev. Lett. 3 (1959) 181.\\

14. G. Goldhaber et al., Phys. Rev. 120 (1960) 300. \\












\end{document}